% TEX SET-UP FOR PROCEEDINGS OF "ASI MEETING HELD AT AHMEDABAD"
% TO BE PUBLISHED IN THE BULLETIN OF ASTRONOMICAL SOCIETY OF INDIA
% Please do not change the page layout, except \hoffset and/or
% \voffset, where changes may be required depending on default printer
% settings. Leave a margin of 3.8 cm on the left and 5 cm at the top
\input psfig.sty
\magnification=\magstep0
\hsize=13.5 cm               %  horizontal size of printed page
\vsize=19.0 cm               %  vertical size of printed page
\baselineskip=12 pt plus 1 pt minus 1 pt  % The line spacing
\parindent=0.5 cm  % The paragraph indent
\hoffset=1.3 cm      % The horizontal offset (may need to be changed)
\voffset=2.5 cm      % The vertical offset (may need to be changed)
\font\twelvebf=cmbx10 at 12truept % Set bold font for Title
\font\twelverm=cmr10 at 12truept % Set large font for Name
\overfullrule=0pt
\nopagenumbers    %  Actual page nos. will be inserted by the Editor
%
% The headlines
% The changes in the headlines should be made just before the Abstract
\newtoks\leftheadline \leftheadline={\hfill {\eightit Saha et al.} 
\hfill}
\newtoks\rightheadline \rightheadline={\hfill {\eightit 
Speckle interferometric observations of close binary stars}
 \hfill}
% Do not change the headline on the first page of paper.
\newtoks\firstheadline \firstheadline={{\eightrm Bull. Astron. Soc.
India (2002) {\eightbf 30} } \hfill}
\def\makeheadline{\vbox to 0pt{\vskip -22.5pt
\line{\vbox to 8.5 pt{}\ifnum\pageno=1\the\firstheadline\else%
\ifodd\pageno\the\rightheadline\else%
\the\leftheadline\fi\fi}\vss}\nointerlineskip}
%
% Defining 8-pt fonts for figure captions and references
\font\eightrm=cmr8  \font\eighti=cmmi8  \font\eightsy=cmsy8
\font\eightbf=cmbx8 \font\eighttt=cmtt8 \font\eightit=cmti8
\font\eightsl=cmsl8
\font\sixrm=cmr6    \font\sixi=cmmi6    \font\sixsy=cmsy6
\font\sixbf=cmbx6
%
%for switching to eight point type \eightpoint
\def\eightpoint{\def\rm{\fam0\eightrm}
\textfont0=\eightrm \scriptfont0=\sixrm \scriptscriptfont0=\fiverm
\textfont1=\eighti  \scriptfont1=\sixi  \scriptscriptfont1=\fivei
\textfont2=\eightsy \scriptfont2=\sixsy \scriptscriptfont2=\fivesy
\textfont3=\tenex   \scriptfont3=\tenex \scriptscriptfont3=\tenex
\textfont\itfam=\eightit  \def\it{\fam\itfam\eightit}%
\textfont\slfam=\eightsl  \def\sl{\fam\slfam\eightsl}%
\textfont\ttfam=\eighttt  \def\tt{\fam\ttfam\eighttt}%
\textfont\bffam=\eightbf  \scriptfont\bffam=\sixbf
\scriptscriptfont\bffam=\fivebf \def\bf{\fam\bffam\eightbf}%
\normalbaselineskip=10pt plus 0.1 pt minus 0.1 pt
\normalbaselines
\abovedisplayskip=10pt plus 2.4pt minus 7pt
\belowdisplayskip=10pt plus 2.4pt minus 7pt
\belowdisplayshortskip=5.6pt plus 2.4pt minus 3.2pt \rm}
%
% define the displayed equations to be indented 1.5 cm from left
% as required by the Bulletin. Hopefully this will work for all
% equations. With this definition using the normal $$...$$ should
% produce the equations with correct indentation, but it will be necessary
% to use \eqno to put equation numbers (though it will be possible to put
% blank eq. nos.). Further, eq. nos using \eqalignno will not work
%
\def\leftdisplay#1\eqno#2$${\line{\indent\indent\indent%
$\displaystyle{#1}$\hfil #2}$$}
\everydisplay{\leftdisplay}
%
% Some useful definitions
% less than or order of \la
\def\frac#1#2{{#1\over#2}}

% greater than or order of \ga

%
%
%to generate boldface characters
\def\pmb#1{\setbox0=\hbox{$#1$}\kern-0.015em\copy0\kern-\wd0%
\kern0.03em\copy0\kern-\wd0\kern-0.015em\raise0.03em\box0}
%
%Beginning of Document%
\pageno=1
\vglue 50 pt  %Leave some space on page 1 before the title
% The title
%
\leftline{\twelvebf Speckle interferometric observations of close binary stars}
% if more than one line is required for the title, then use next two lines ...
%
\smallskip
%\leftline{\twelvebf Proceedings of ``ASI meeting held at Ahmedabad''}
% end of title
\vskip 40 pt  % Space between title and author(s) name(s).
\leftline{\twelverm S. K. Saha$^1$, V. Chinnappan$^1$, L. Yeswanth 
\footnote{$^1$}{\eightit e-mail: sks@iiap.ernet.in, vchin@iiap.ernet.in, 
yeswanth@iiap.ernet.in, \ \ $^2$anbu@iiap.ernet.in},
and
P. Anbazhagan$^2$}
% Name of Authors
\vskip 4 pt
\leftline{\eightit $^1$Indian Institute of Astrophysics, Bangalore 560 034. 
India.}
\leftline{\eightit $^2$ Vainu Bappu Observatory, Kavalur 635 704.
India.}
%\leftline{\eightit  to reduce the number of lines}
%
% If authors are from different institutes, repeat the above lines
% for each institution. For authors from same institution write the
% names in one line.
%
%\vskip 0.5 cm
%\leftline{\twelverm V. R. Co-author1 and V. R. Co-author2}
%\vskip 4 pt
%\leftline{\eightit Name and Address of the institution}
\vskip 20 pt % leave some space between author(s) names(s) and abstract
%
%
% The leftheadline should include the Authors' name, for two authors use
% \&  (e.g. I. M. Author \& I. M. Co-author) for three or more authors
% use et al.,
\leftheadline={\hfill {\eightit Saha et al.} \hfill}
% Use a short running title as the rightheadline
\rightheadline={\hfill {\eightit Speckle interferometric observations of close 
binary stars} \hfill}

% Abstract begins
%
{\parindent=0cm\leftskip=1.5 cm

{\bf Abstract.}
\noindent
Speckle interferometric technique is employed to record a series of hundreds of
short-exposure images of several close binary stars with sub-arcsecond
separation through a narrow band filter at the Cassegrain focus of the 2.34 
meter (m) Vainu Bappu telescope (VBT), situated at Vainu Bappu Observatory 
(VBO), Kavalur, India. The data are recorded sequentially by a Peltier-cooled 
intensified CCD camera with 10~ms exposure. The auto-correlation method 
is applied to determine the angular separations and position angles of these 
binary systems.
\smallskip 
\vskip 0.5 cm  %  Space between Abstract and Key words
{\it Key words:} interferometer, speckle imaging, close binary stars.

}                                 %  End of abstract
% Beginning of document
%
%
% Beginning of a section heading
%
% for the first section leave 20 pt space, for subsequent sections just
% leave bigskip (i.e. 12 pt)
\vskip 20 pt
\centerline{\bf 1. Introduction}
\bigskip
\noindent  
Speckle interferometric technique (Labeyrie, 1970) is widely used
to decode the deleterious effect of the atmospheric turbulence that limits
the resolution of ground-based telescopes. Recent reviews by Saha (1999, 2002)
describe in depth the utility of this method. One of the important subjects in 
stellar astrophysics is the studies of close binary stars. This interferometric 
technique deciphers the diffraction-limited autocorrelation of the object. 
In case of the components in a group of stars, this method retrieves the 
separation, position angle with 180$^\circ$ ambiguity, and the relative 
magnitude difference at low light levels. This paper presents
the results obtained at the Cassegrain end of the 2.34 m VBT, VBO, Kavalur using
such technique. Data analysis was carried out using the algorithm developed by
Saha and Maitra (2001).
\vskip 20 pt
\centerline{\bf 2. Instrumentation \& Observations}
\bigskip
\noindent
Large number of speckle-grams of several close binary stars and of 
reference stars were recorded on 10th. April, 1999 using the speckle 
interferometer (Saha et al., 1999) at the Cassegrain 
focus of the VBT through a 10~nm filter centered on 6761~\AA. These images were
acquired with a Peltier-cooled ICCD camera with an exposure
time of 10~ms. This camera offers options of
choosing the exposure time, viz., 1~ms, 5~ms, 10~ms, 20~ms etc. It can operate
in full frame, frame transfer and kinetic modes.
Since the CCD is cooled to - 40$^\circ$C, the dark noise is low. Data are
digitized to 12 bits and can be archived to a Pentium based PC.
In full frame, as well as in frame transfer modes, the region of interest can
be acquired at a faster speed. While in the kinetic mode, the image 
can be confined to a small area.
\vskip 20 pt
\centerline{\bf 3. Data reduction and the results}
\bigskip 
\noindent
Data of two close binary stars, HR4891 and HR5298 alongwith the respective 
reference stars were processed. The data acquired using window 3.1 software is 
stored in the SPE format. The data is then converted to FITS (16 bit unsigned 
integer) format. The SPE format has a fixed header size of 4100 bytes, followed 
by an array of 16 bit integer values, the size of which is determined by the 
image section used and the number of frames acquired in a single sequence.
\bigskip 
\noindent
More than 400 frames of each of the aforementioned stars are scanned
carefully. Some of the frames contained specklegrams whose
centroid were displaced with respect to the circular pinhole image.
In the case of binary stars, these images considerably reduce the visibility of 
the fringes in the processed image. Since the Fourier transform method of image 
reconstruction is effective only when the noise content in the acquired image is
within acceptable limits, each frame was inspected, and the frames in which the 
centroid of the specklegram was centered within the pinhole image to within 
10-15 pixels were selected. The elimination of the faulty frames in this manner 
has improved the fringe contrast by a factor of 2. 
The separation of the components of HR4891 and HR5298 are found to be 
0.094$^{\prime\prime}$ and 0.109$^{\prime\prime}$ respectively. These results 
are consistent with the CHARA catalogue. Figure 1 depicts the autocorrelation of
the binary star HR5298; the companion star is imprinted on either side
of the primary star.
\bigskip
\noindent
% inserting figures.
\midinsert
%Leave appropriate space to paste the figure
%\vskip 6.0cm
{\eightpoint   % Switch to 8 pt fonts for figure captions
\noindent
\centerline{\psfig{figure=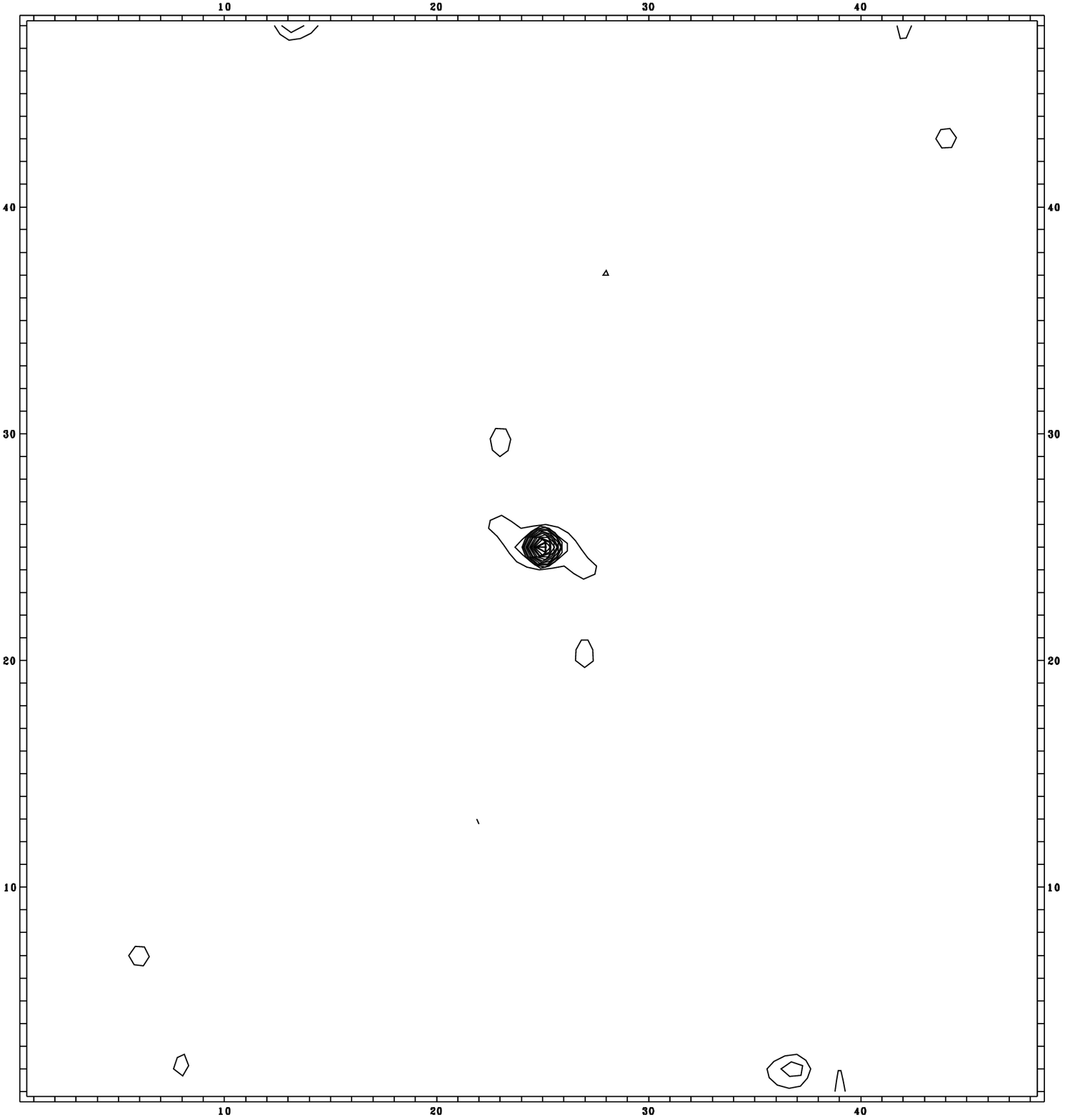,height=7cm,angle=270}}
{\bf Figure 1. Autocorrelation of HR5298.}
}
%  End of 8 pt fonts, leave one line blank after the caption
%        to set the caption with reduced baselineskip for 8-pt font.
\endinsert
\centerline{\bf References}
\bigskip
{\eightpoint\parindent=0pt\everypar={\hangindent=0.5 cm}
% References in the format of the Bulletin of the Astronomical Society of India
% using 8 pt fonts
% leave one line blank between two references to force a paragraph break
%scussions.

CHARA third Catalogue, 1997, eds. W. Hartkopf, H. McAlister, and B. Mason.
 
Labeyrie A., 1970, A \& A, 6, 85.

Saha S. K., 1999, BASI, 27, 443.

Saha S. K., 2002, Rev. Mod. Phys., (April issue).

Saha S. K., Maitra D., 2001, Ind. J. Phys. 75B, 391.

Saha S. K., Sudheendra G., Umesh Chandra A., Chinnappan V., 1999,
Experimental Astronomy, 9, 39.

% End of section heading
%\endref
}                                         % End of references
% leave one line blank before the closing braces.
\end